# Lists of Top Artists to Watch computed algorithmically

Tomasz Imielinski

Articker.org

**Abstract**

Lists of top artists "to watch" are periodically published by various art world media publications. These lists are selected editorially and reflect the subjective opinions of their creators.

We show an application of ranking by momentum method [1] to algorithmically produce lists of artists to watch derived from our platform Articker [2]. We use our algorithms every month to produce a new list of momentum leaders on www.articker.org.

The lists of momentum leaders computed this way has the following properties:

  a) It is small (because of the small frontier property [1]).

  b) It is unbiased – with no bias towards famous artists nor emerging, and yet unknown, artists

  c) It is fair – artists who are not included in the list must be Pareto dominated by at least one member of the momentum leaders list (Pareto frontier)

  d) It is objective – it is computed automatically, not editorially selected.

## 1. Introduction

There are many *"Top artists to watch"* lists published every year. Some examples include [3]-[6]. These lists are editorially created and obviously, they are subjective. For example [6] shows the artists to watch list produced by Whitewall magazine in 2020, which consists of such artists as *Matthew Wong, Amoako Boafo, Titus Kaphar, Mequitta Ahuja, Petrit Halilaj, Genieve Figgis. Ida Ekblad. Salman Toor, Tala Madani, Shara Hughes, and Paul Mpagi Sepuya.* These are excellent choices, which are purely editorial and some may be controversial. For example, one may argue why neither *Julie Curtiss nor Tschabalala Self* is on this list. But this is the nature of editorial choices which reflect the taste of the curators of such a list. They do not have to explain why some artists have been omitted.

Imagine however such a list of Top artists to watch, where a clear explanation for being on the list, as well as *not* being on it, can be provided.

This is what we propose in this paper.

Specifically, we propose an objective - algorithmically created "Top artists to watch" lists based on two dimensions. Our momentum driven lists are based on the absolute gain and relative gain of the media presence of artists reflected by Articker's media index [2]. We produce such lists monthly on articker.org. We use the ranking by momentum method proposed in [1] and apply it to different windows of time, using Articker's media index to produce absolute and relative gains. Ranking by momentum uses Pareto ordering over two dimensional vectors of absolute and relative gain. Ranking in two dimensions allows a balance between famous and emerging artists. Indeed, ranking only by absolute gain would favor famous artists. Ranking only by relative gain would promote unknown, emerging artists with low media index values.

Our lists are created with transparency by an easy-to-understand algorithm based on data harvested from the public domain. Not only do we justify the presence of an artist on the list, but we also, for each artist who did not make it, answer a fundamental question – why the artist is not on the list.

**REMARK:** *There are some artists lists which are ranked in one dimension, such as number of Instagram views, total auction turnover last year, number of searches on artsy etc. These lists are typically biased towards already famous artists. We could use any of these metrics: Instagram views, auction turnover, number of searches as the basis for our momentum driven ranking method: it would be momentum of Instagram views, momentum of auction turnover or momentum of number of artist searches on a web site. Our method applies to any artist score - and by no means it is limited to the media index.*

Here are several recent lists on articker.org, the top one from November 2021, followed by October and September of 2021, back to June 2021.

**Articker**

TECHNOLOGY    SPOTLIGHTS    PAPERS    PODCASTS    IN THE NEWS    ABOUT US

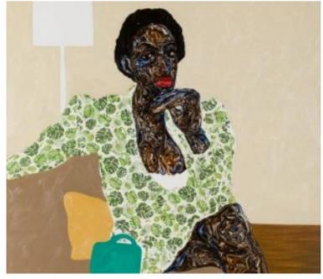

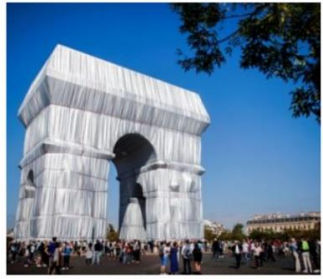

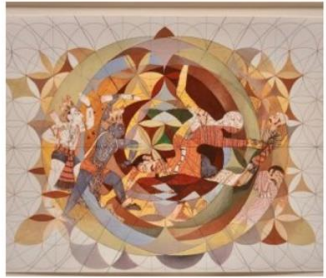

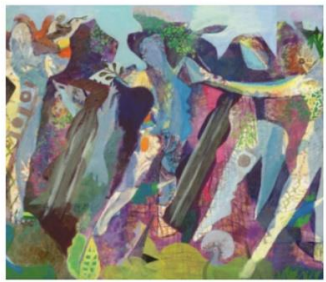

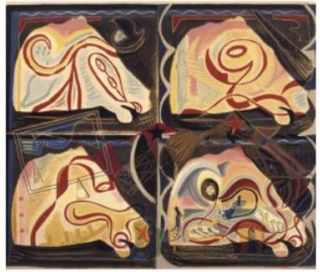

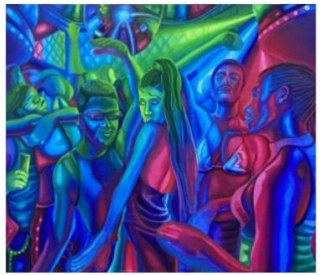

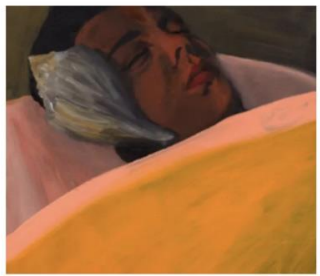

**Figure 1.** *Momentum leaders lists on www.articker.org.*

A fragment of one of our lists is presented in Figure 2. Artists are ranked by Momentum, computed by our method from Articker media index data. For each artist, we list their Media index as well as the value of weighted momentum. In addition, three links to the recent important new articles along with the pictures of the artist's work extracted from each article.

## Rank #1: Sophie Taeuber-Arp

Media-Index: 3021
Momentum: 28.06%
Featured In:
Art&object: **Sophie Taeuber-Arp Receives Spotlight at Tate Modern**
The Times: **Sophie Taeuber-Arp at Tate Modern — the forgotten genius of artists' wives**
Wallpaper: **Haegue Yang on the legacy of Sophie Taeuber-Arp**

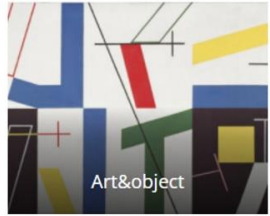 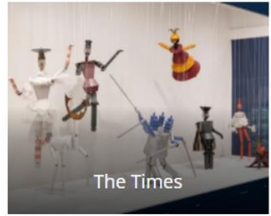 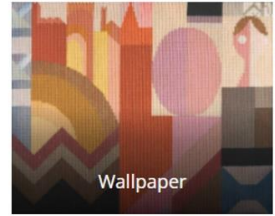

## Rank #2: Paula Rego

Media-Index: 8605
Momentum: 25.7%
Featured In:
The Arts Desk: **Paula Rego, Tate Britain review – the artist's inner landscape like never before**
The New Yorker: **The Fury and Mischief of Paula Rego**
The Guardian: **Paula Rego review – stunning is an understatement**

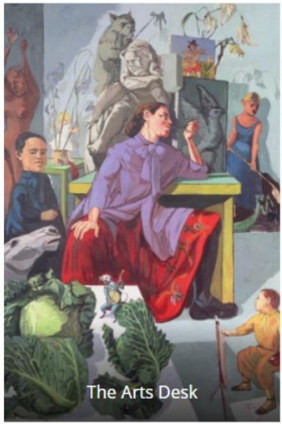 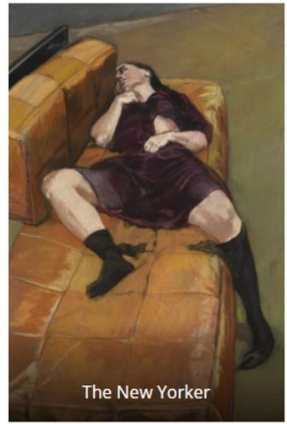 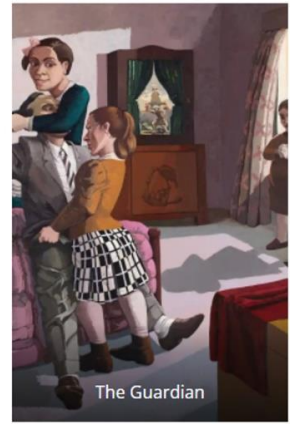

## Rank #3: Eileen Agar *

Media-Index: 1734
Momentum: 19.22%
Featured In:
Juxtapoz: **Eileen Agar: Angel of Anarchy @ Whitechapel Gallery, London**
Apollo Magazine: **For Eileen Agar, the natural world was a playground of artistic possibilities**
Art mag: **An Anarchic Angel – Eileen Agar at Whitechapel Gallery, London**

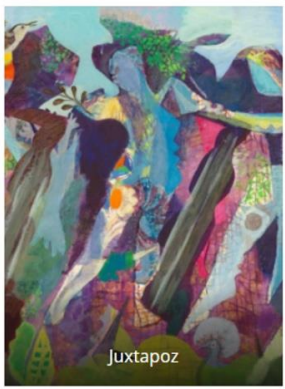 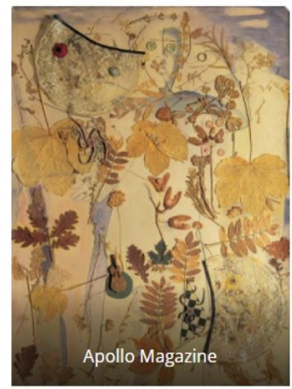 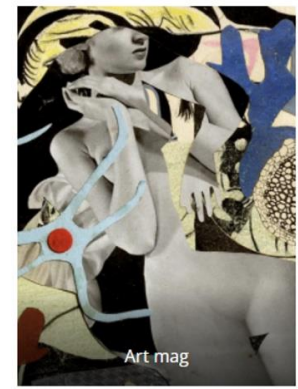

**Figure 2.** *Example of top 3 artists on one of the articker.org lists.*

How are our lists of momentum leaders constructed?

We will begin with recalling the basic definitions from [1].

Let E be a linearly ranked set of N entities and T be a time window. We will refer to the entities by their ordinals ranging from 1,...N, where 1 refers to the entity ranked first and N to the one ranked last. Ranking can be based on some underlying scoring function with entities ranked in descending order of the score. But the absolute value of the score is irrelevant here – just the ranking that it imposes. For example, YouTube videos can be ranked by the total number of views, Stocks by market capitalization, artists by the Articker media index, etc.

Let g: <1...N> -> R be a function called *absolute gain* (over period T)   Let r be another  function from <1...N> ->R called *relative gain* (over period T).

The triple

E*=<N, g, r>

Is called a Δ-system.

Δ-system represent a single snapshot of change of all entities from 1 to N, over the same period of time T. The period of time T can range from seconds to months. The absolute gain and relative gain functions refer to changes of entity scores over T. Absolute gain g refer to the absolute gain of score over period T. Relative gain,  r  refers to percentile points –  comparison of absolute gain to the score of the entity before the period T).

Our entities are artists (about 100,000 of them) and absolute gain and relative gain refer to the media index gain over 90 days. We define the media index of an artist further below.

Given two entities e, f, we define

$e <<_m f$  iff   $g(e) < g(f)$ and $r(e) < r(f)$

Clearly $<<_m$  is Pareto ordering, the partial order, which is not total. That is, there are incomparable pairs of entities e and f, such that neither

$e <<_m f$ nor  $f <<_m e$ is true

By a *momentum leader* in the  Δ-system  <N,g,r>, we mean a maximal element in $<<_m$. The set of momentum leaders is the Pareto frontier of the Pareto ordering $<<_m$

Ranking by momentum [1] was motivated by our work on Articker and we would like to describe here how ranking by momentum can be used for automatically establishing artists "to watch" based on the momentum over a window of time (90 days in our case).  Ranking by momentum can be applied to the absolute and relative gain of any score – from the market capitalization of stock to the number of views of a YouTube video.  In this paper, the Articker's media index of an artist is the underlying score of our momentum ranking and its momentum leaders.

**Media Index**

Articker (articker.org), measures the media presence of over 100,000 artists in nearly 16,000 media sources through ***media index***. The media index reflects the cumulative presence of an artist in all online publications until a given time t.  It is a refinement of just number of mentions of the artist's name in all online publications.  Every time an artist's name is mentioned in an online publication,  their media index is updated. In addition, the media index depends on the domain which published the article as well as by the position of the artist name in the article itself and several other parameters. For example publication in New York Times carries a much higher domain weight than publication in a University Art Blog. Similarly, an artist's name appearing in a headline of an article carries much more weight than just mention of the name in the article's body. Longer articles matter more. Articles with fewer artists mentioned, also count for more in the media index of a given artist. In particular,  even when the artist name is not mentioned in the headline,  but is the only artist name mentioned in the article's content, such article is likely to be a feature about this artist and as such should carry more weight.

Articker system continuously calculates the media index for each artist every couple of hours, every day, week, and month of each year.

The media index serves as an underlying score to calculate absolute and relative gains necessary to form Pareto ordering.  One could also use a simple mention count as the underlying score and form Pareto ordering based on the absolute and relative gain of just number of artist mentions. We decided to use the Media index since it is simply a more sophisticated measure of artist presence in the media.

Articker users who are either art collectors or art specialists want to monitor who from 100,000 plus artists were uptrending the most, who carried the most momentum?  Who stood out from the incomprehensively large cohort of artists, last summer, Last year, even last month.

Every month we use ranking by momentum to publish a list of 8-12 artists who are momentum leaders over the period of the last 90 days.  These are based on Pareto ordering of two-dimensional vectors – of absolute gain and media share gain/loss of each artist over the preceding 90 days.  It is important to point out that we publish the whole Pareto frontier, i.e. the whole set of momentum leaders.  Thus, there is no editorial intervention in the choice of momentum leaders.  As demonstrated experimentally in [1], when absolute and relative gain follow power-law, the Pareto frontier is small (of the order of O(log(N)^2)). This is how we almost invariably get a list of between six and fifteen momentum leaders.

Such lists have been published since June 2021, with the latest one published in mid-October, 2021.

One of the main properties of our algorithm is that for each artist who is NOT INCLUDED we have justification for his/her non-inclusion. It is because s/he has been Pareto dominated by a specific artist who is among our momentum leaders.

For example, on August 15th we have published https://articker.org/spotlights/top-10-artists-with-most-momentum-last-90-days/ the list of top 10 artists by momentum in the last 90 days. The top 10 by momentum list was led by geometric artist Sophie Taeuber-Arp (d. 1943) and followed by Paula Rego both gaining momentum due to enthusiastically received recent Tate Museum exhibitions and Eileen Agar a surrealist (d 199) who 30 years after her death is experiencing a comeback.

We first started by simply ranking artists by media share gain (this is a variant of relative gain as generally defined earlier in this paper). This was our first list from July 2021. Unfortunately, such ranking had a serious shortcoming - it was heavily biased towards less known artists, with smaller media indexes. This did not seem fair at all since it would shun better-known artists who in fact may be experiencing a major uptrend. These observations led us to the idea of ranking by momentum [1], which applies much more widely than just for Articker. In this paper, we describe how ranking by momentum is used in determining our monthly lists. We also show various properties of our lists as well as even stricter small frontier property which has been satisfied for 70 potential lists over windows of time ranging from 90 days during the period since 2015.

## 2. Articker Data

Let us summarize our data first. We will show media index value distributions as well as relative gain distributions among artists in our database.

The plot below shows the power-law distribution of the media index values for around 100,000 artists. We can see that the scores range from nearly 500,000 (Picasso) to fractions of less than 1. The median media index is around 35, therefore half of all artists have media indexes of less than 35

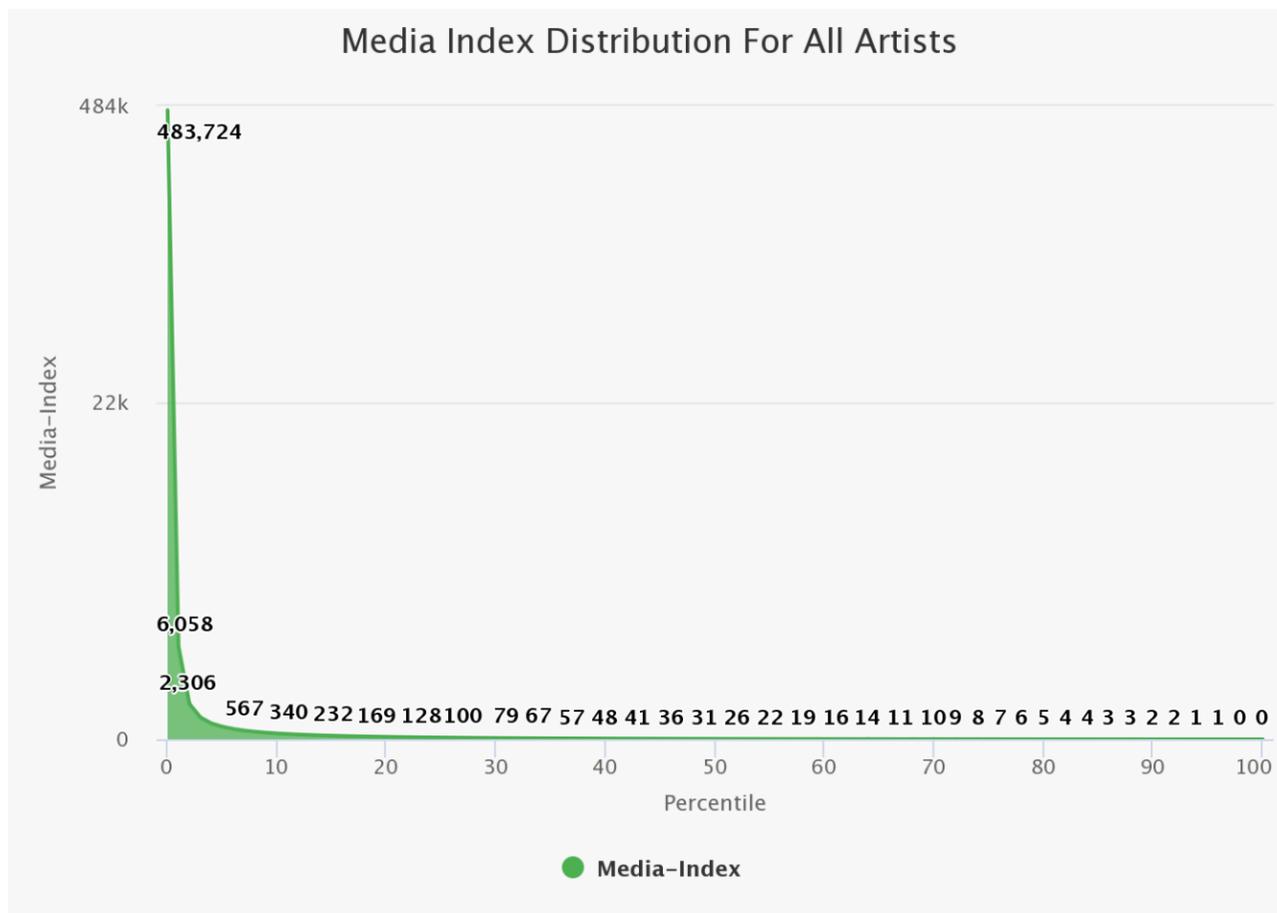

Figure 3. *Power-law distribution of Articker's media index scores.*

The media share of an artist is defined as their media index divided by the total media index of all artists. While the media index is a monotonically increasing function of time, media share can increase or decrease. Increasing media index share indicates that artist is gaining media index faster than the "market" – the set of all artists. A large gain in media index share (measured usually in percentage points) indicates an up-trending artist.

Media share gains are generally much higher for artists with lower media index values. It is effectively a form of relative gain, we described in the earlier part of the paper.

In general media share gains for artists are distributed according to power law as well, even for subsets of artists sharing similar media index values.

For example, media share gains for artists with a media index of around 500 are distributed according to a steep power-law distribution. The distribution ranges from nearly 300% to negative values (losses).

The plot below shows only artists who are gaining media index share. This is why, in this plot, all values of media index gain are positive. If we made a similar graph for artists with a score around 50000, the maximal gain would generally be no more than 10%. The larger the media index of a cohort of artists, the smaller the media share gain values. But distributions would still follow power-law – just scaled down for artists with the higher media index value.

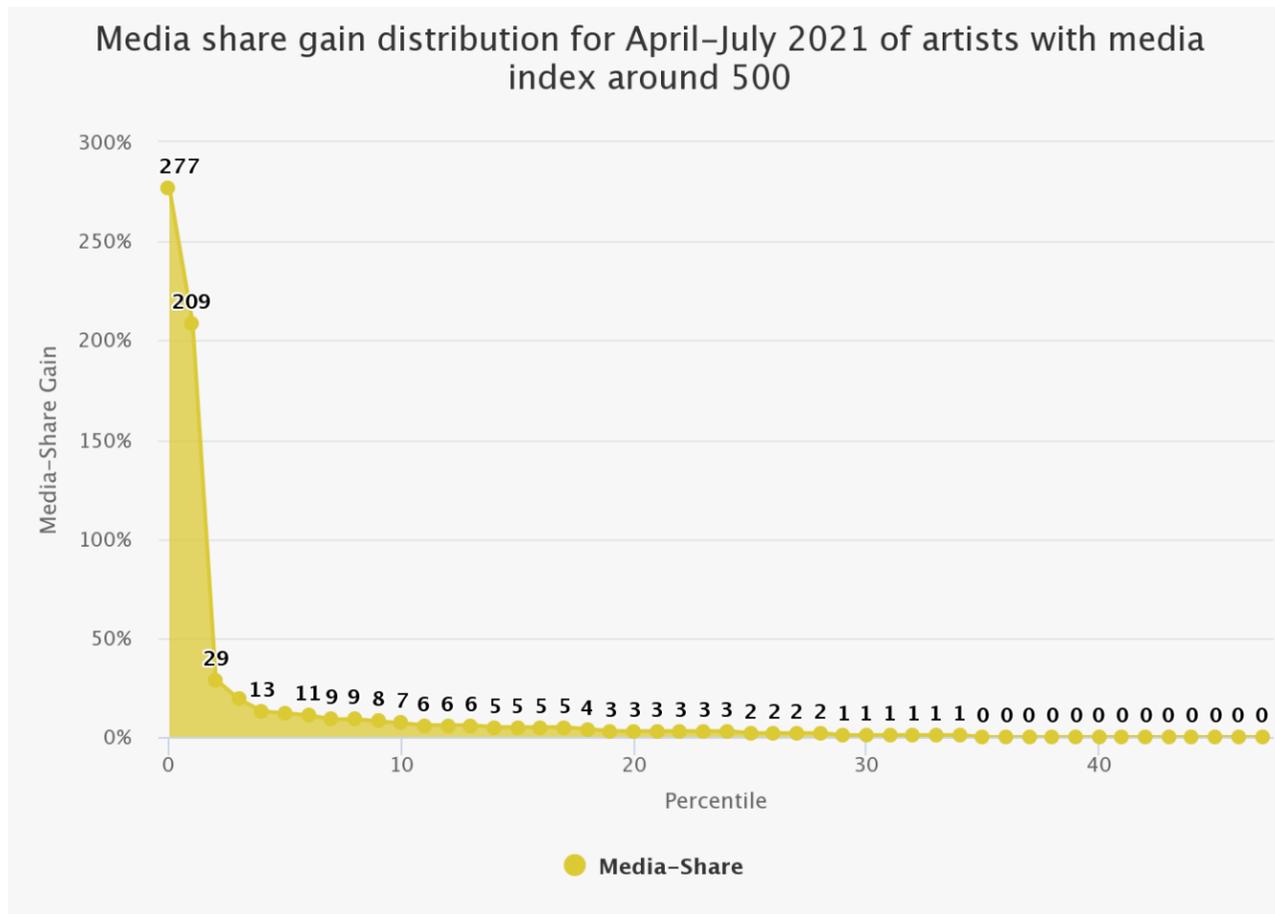

Figure 4: *Power-law distribution of Media Share gains for a cohort of artists.*

**Inversion between Relative and Absolute Gain**

We have examined around 20,000 artists who have significant media coverage. We have ranked them by the total media index score which reflects their lifetime media coverage.

The plots below show the inverse relationship between the Relative Gain and Absolute Gain.

The first two plots shows:

RelativeGain(k) = Min{i: r(i) > $2^k$} for Relative gain and

AbsoluteGain(k)= Max{i: a(i) >$2^k$} for Absolute gain respectively.

We can observe that there are two areas of very steep ascend (Relative Gain) and steep decline (Absolute Gain). For example, for k=5,6,7 the minimum rank of an artist with a relative gain larger than 32, 64, and 128 is 5100, 11,400, and 18,300 respectively. Similarly for the k=6, 7, 8 the maximum rank of artists with absolute gain larger than 64, 128, and 256 respectively are 18,500, 11,200, and 5000+, declining exponentially in a similar region to an exponential increase of the Relative Gain graph.

This is intuitive since relative gain values generally increase with lower-ranked artists whose media index scores are low. Similarly, the absolute gain values increase with higher-ranked artists whose media index scores are high.

Not only do absolute gain and relative gain behave "inversely" but also both obey, in the mid-section of their graphs, power-law with exponent larger than 1. These distributions are responsible for the small frontier property of Articker data as evident through the analysis below. They also confirm generally the Small Frontier Property from [1].

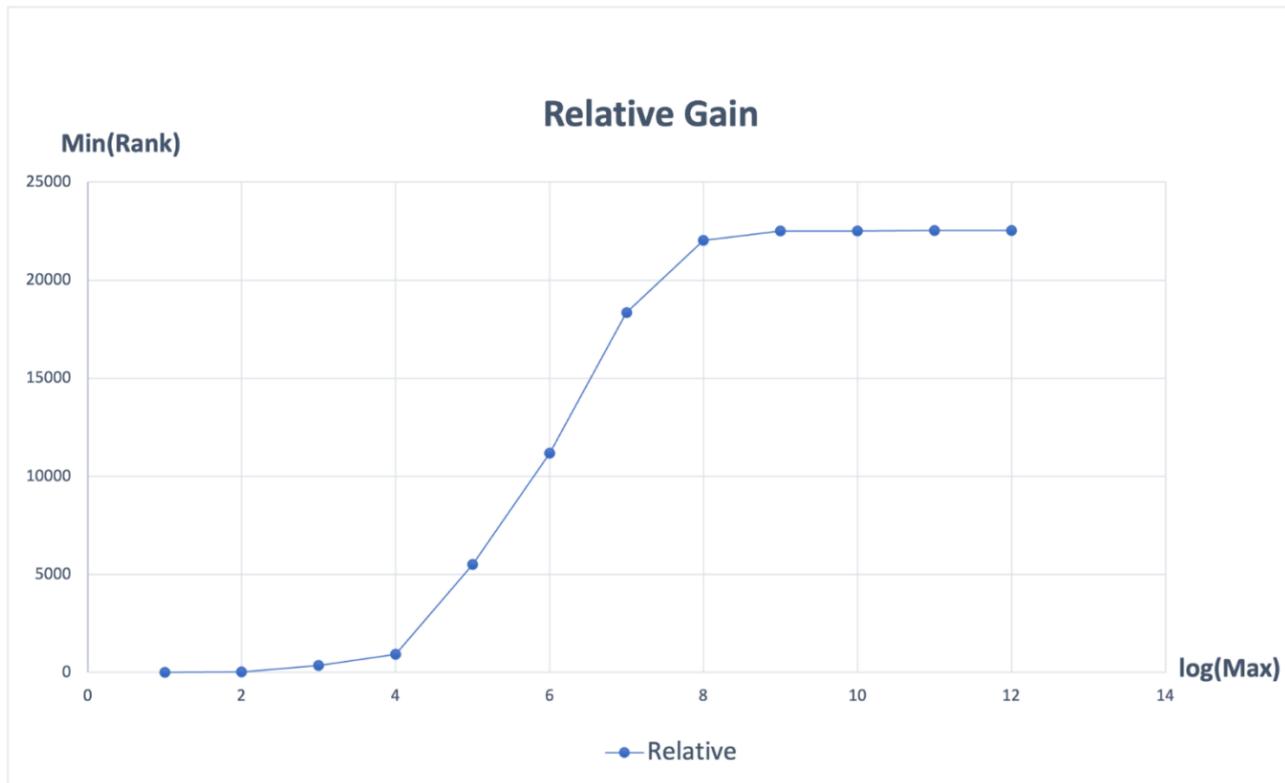

**Figure 5.** *Minimum media index rank of artists exceeding a relative gain of X (logarithmic scale).*

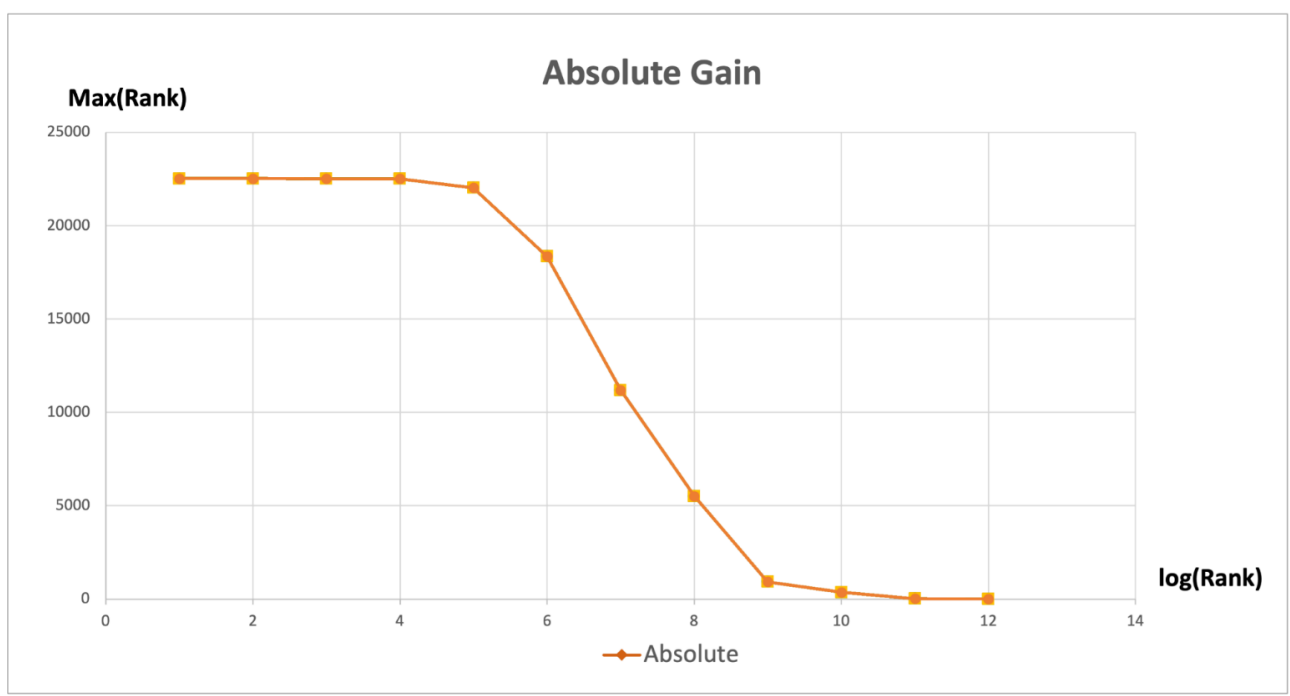

**Figure 6.** *Maximum media index rank of artists below the total gain of X (logarithmic scale).*

Finally, we should the superpositions of both graphs to show the almost perfect inverse symmetry.

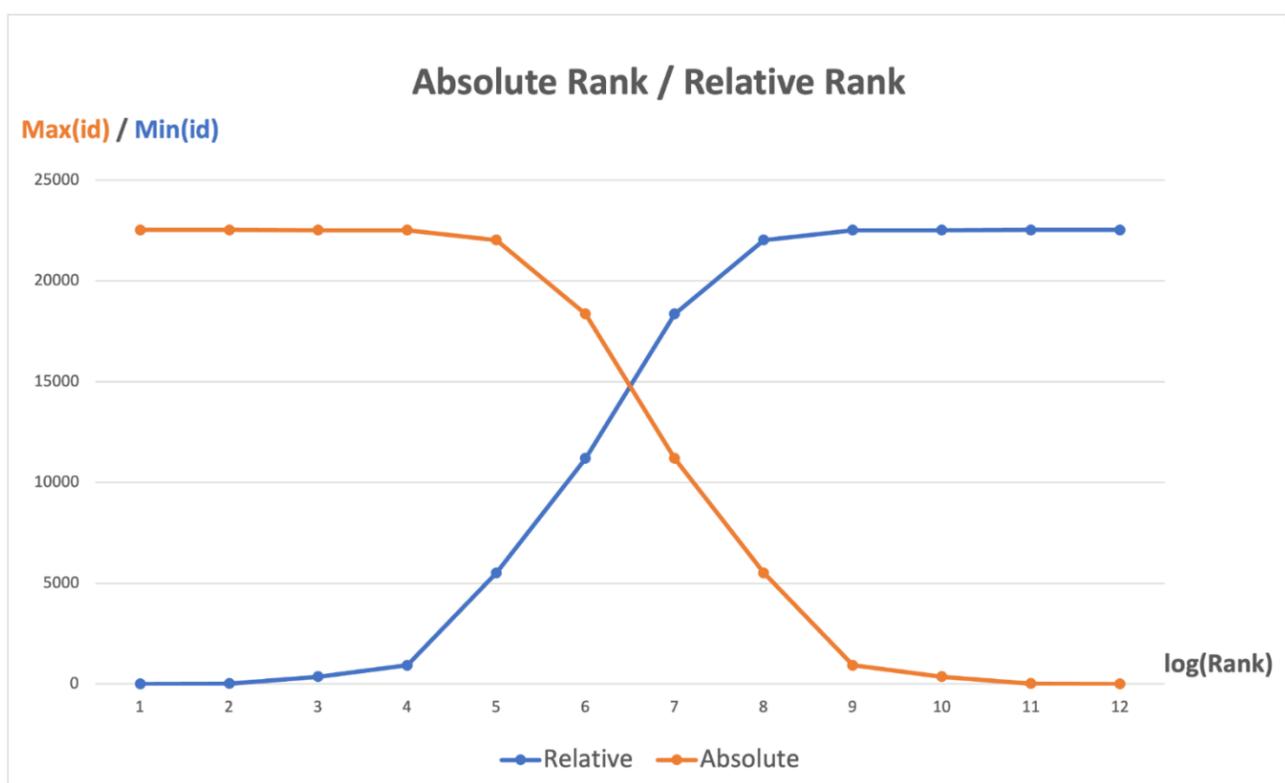

**Figure 7.** *Both graphs (Figure 5 and 6) are superimposed.*

We have examined 70 windows of 90 days each, ending in months ranging from January 2015 till October 2021. For each such window, we have determined the size of the Pareto frontier – the number of momentum leaders. The results are strongly consistent with the small frontier property. The average size of the Pareto frontier is slightly less than 10. This is the usual size of editorially formed lists of top artists to watch. But computed by an algorithm and guaranteeing that any artists among 20,000+ are dominated in Pareto ordering by one of the 8-15 momentum leaders in each of the 70 cases.

| Size of Pareto Frontier | Number of 90-Day Windows |
|---|---|
| 6-9 | 24 |
| 9-12 | 36 |
| 12-15 | 10 |

**Table 1.** *Distribution of sizes of Pareto Frontier for past windows of 90 days since January 2015.*

### 3. Top K views and Momentum Leaders

Before we dive deeper into the specific list of momentum leaders on August 1, 2021, we would like to offer another view of our data.

We have taken the distribution of rankings of the top 10, 50, 100, and 500 artists according to absolute gain as well as relative gain (media share gain) over the last 3 months. For this analysis, we have taken one of the current windows, the one which ended on August 1, 2021, and covered the preceding 90 days.

The following multi boxplot shows the distribution of 90% of highest ranked members of each group – top 10, top 50, top 100, and top 500 by total gain and by relative gain.

For example, the first box shows the distribution of the top 10 artists by the total gain in the last 3 months is. 90% (i.e. 9 artists) of the top 10 artists by total gain are ranked between 1 and 11 by overall score. This reflects the bias of total gain ranking towards the top artists such as number one Picasso, number two Warhol, etc. The average ranking is 9.3 (which is affected by the outlier in this group, ranked 45, Joseph Beuys).

The next box, colored blue, reflects the distribution of rankings of the top 10 artists by relative gain. We see this box much higher than the leftmost box – reflecting a bias towards low-ranked artists. The second row shows that such artists (90% of them) populate the interval between 15,642 rank and 19865 ranks, with an average rank of 17017 (marked red) – very much the bottom of the overall ranking.

The next pair of boxes corresponds to the top 50 artists by absolute gain and relative gain. The isolated points show the outliers. For example, we see that artists ranked 2169 and 6510 belong to the top 50 relative gainers, although they are ranked way much higher (lower ordinal) than "their" blue box. Similarly, the artists ranked 326, made the list of top 50 gainers despite being ranked much lower than the black box upper limit of 161. Other outliers include ranks of 2169 also for both the top 100 in absolute and relative gain ordering (black box and blue box) as well as rank 1183 for the black box. Finally, the rank 7477 is a far outlier for the top 500 black box (upper bound of 2845) and 1183 is a very far outlier of the last blue box with a lower bound of 7360.

Who are these outliers?

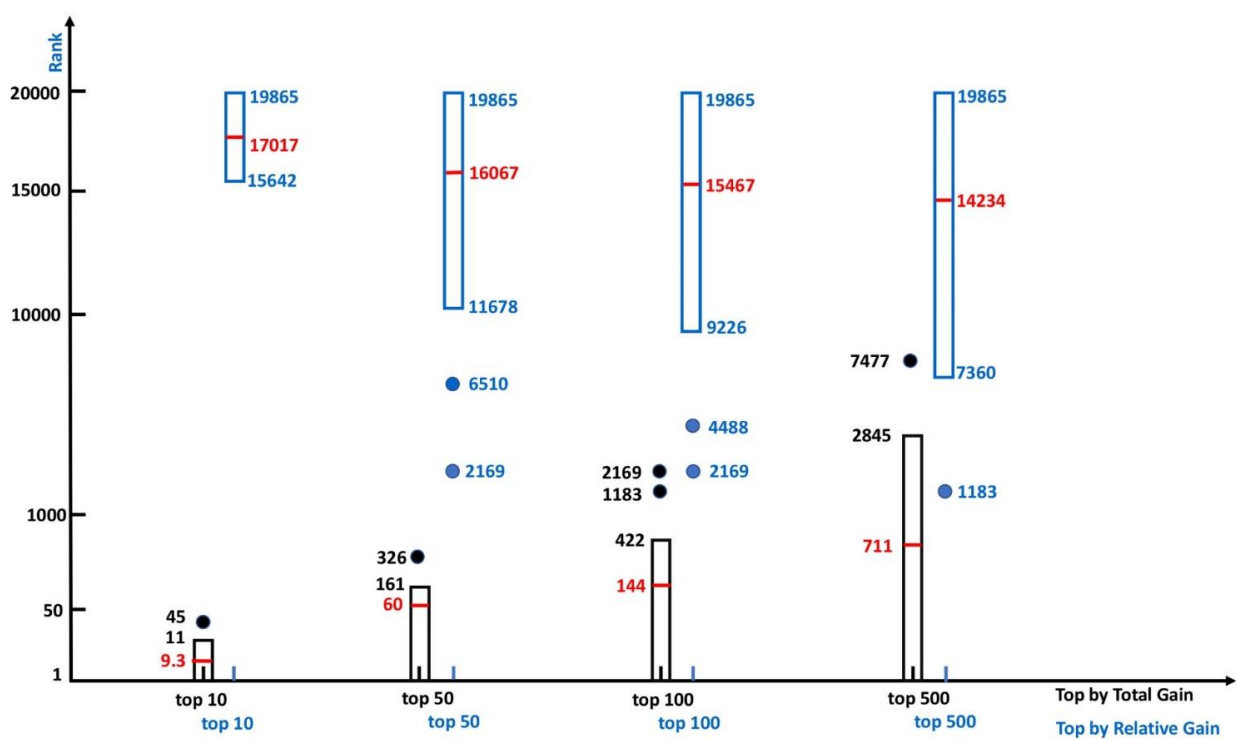

**Figure 8.** *Box plots for top K list for absolute and relative gains respectively. Each box shows the maximum (minimum) rank of the 90th percentile of the top K list. Outliers are marker by blue and black circles.*

**Who are the outliers? Who is "out of the box"?**

Outliers are most interesting since they owe their status to the combination of total gain and relative gain. Thus, as we will see, they top the list of momentum leaders.

So who are the artists ranked 45[th], 326[th], 2169[th], 1183[th], and 7477[th]? Turns out that except one, they all momentum leaders and members of Pareto frontier of our ordering $>>_m$

Before we disclose their names, let us show the ranked list of momentum leaders for the August 1 order.

This is the list as of August 1, which predates the articker.org list by about two weeks. As we can see our momentum ranking is very dynamic, and some artists move up (like Paula Rego) due to the most recent intensity of news – in the two weeks between august 1 and august 15.

| Id | Name | Momentum | IntervalL | IntervalR | Totalgain | Relativegain |
|---|---|---|---|---|---|---|
| 2169 | Eileen Agar | 24.77 | 1183 | 6510 | 801.55 | 80.6 |
| 1183 | Sophie Taeuber-Arp | 22.34 | 415 | 2169 | 866.66 | 35.62 |
| 45 | Joseph Beuys | 18.7 | 11 | 194 | 3270.62 | 6.49 |
| 7399 | Danielle Mckinney | 16.75 | 4486 | 15642 | 332.62 | 198.52 |
| 415 | Maya Lin | 12.25 | 326 | 870 | 943.58 | 11.84 |
| 9 | Basquiat | 10,58 | 3 | 45 | 7124.31 | 4.07 |
| 269 | Alice Neel | 9.49 | 118 | 326 | 1137.05 | 9.69 |
| 326 | Paula Rego | 4.81 | 269 | 415 | 1068.54 | 10.51 |
| 3 | Van Gogh | 2.46 | 1 | 6 | 8640.25 | 0.98 |

**Table 2.** *Articker ranking by Momentum over 90 days between May 1, 2021, and August 1, 2021.*

Many outliers made it to the list of Momentum leaders, the impressive outlier, ranked 2169[th] overall – is **Eileen Agar**, who tops our list in Table 2. Eileen Agar, is in the top 100 artists by total gain, despite being ranked 2169[th] by the media index. Thus, her total gain is unusually high for artists ranked so low by the media index. Also, she is in the top 50 artists by relative gain, thus making to the elite artist in *BOTH rankings*. For the relative gain top 50 list, she is an outlier far above the expected range of **[11678, 16067*   19865].**

She is followed by **Sophie Taeuber-Arp,** ranked 1183[th] overall. Joseph Beuys is another outlier, who is ranked 45[th] overall by the media index. **Paula Rego**, ranked 326[th], is another outlier and is ranked 9[th] in Table 2.

**Danielle McKinney**, ranked 7399[th] by media index, is also an outlier, who despite her low media index ranking, made it to the top 500 by total gain, although another outlier ranked 7477[th] overall  -  **Khari Turner** was even further out of his box, also just a little bit. McKinney has dominated Khari Turner in the Pareto ordering. McKinney's total score of 332.62 is higher than Khari Turner's total gain of 211.74 and her relative gain of 198.52% is higher than Khari Turner's 72.11%.

McKinney is in the top 10 artists by relative gain, despite a much higher ranking than the lower bound of the box for this category, which is 15642 – she is over 8000 ranking positions above that. She is 255[th] in total gain rank

What these outliers have in common is that they do well in both "opposite" rankings – the total gain and the relative gain. Thus, their presence in the top 10 momentum list is not a coincidence. This is exactly what the Pareto order $>>_m$ rewards.

**The momentum ranking rewards artists who are ranked high in BOTH total gain and relative gain ranking**

All 10 artists in table 2 are momentum leaders (that is maximal elements in $>>_m$ ranking). They are ranked by the w(m) – as explained in the first section. The leader **Eileen Agar** is ranked 2169[th] overall but she dominates the interval of all artists ranked between the rank 1183 and 6510, this over 5000 artists (!) in the Pareto ordering $>>$. Her total gain and relative gain are higher than any of these 5000 artists. This clearly shows her momentum. She dominates not just her "cohort", i.e artists who are ranked by media index near her, that is around 2100[th], but the much wider population of artists, who are ranked much higher (up to 1183[th] ) and who are also ranked much lower than her (down to 6510[th]) She dominates these artists in both total gain and relative gain. Her domination in total gain (with a total gain of 801.55) of artists ranked higher by almost 1000 ordinals (up to  1183[th] ) is very impressive. Agar's domination in the relative gain of artists ranked over 4000 ordinals below her (down to 6510[th] ) is perhaps even more significant, with 80.5% of media share gain over 90 days preceding August 1.

**Joseph Beuys, Basquiat**, and **Van Gogh** represent blue-chip artists in table 2. Their relative gains are not impressive, since it is very difficult to achieve higher relative gains for high-scoring artists. Notice that Van Gogh with a total gain of 8640.25 and a relative gain of just 0.98% nevertheless dominates just 4 artists ranked lower than 1 and higher than 6 – but who carry enormous scores. Thus his momentum is still 9th, at 2.46%. Basquiat dominates artists ranked between 3 and 45, with his total gain of 7124.31 and relative gain of 4.07%. This results in a momentum of over 10%.

The table below shows a different ranking of the Momentum leaders table. This is the ordering from the lowest to the highest overall rank.

| Id | Name | Momentum | IntervalL | IntervalR | Totalgain | Relativegain |
|---|---|---|---|---|---|---|
| 7399 | Danielle Mckinney | 16.75 | 4486 | 15642 | 332.62 | 198.52 |
| 2169 | Eileen Agar | 24.77 | 1183 | 6510 | 801.55 | 80.6 |
| 1183 | Sophie Taeuber-Arp | 22.34 | 415 | 2169 | 866.66 | 35.62 |
| 415 | Maya Lin | 12.25 | 326 | 870 | 943.58 | 11.84 |
| 326 | Paula Rego | 4.81 | 269 | 415 | 1068.54 | 10.51 |
| 269 | Alice Neel | 9.49 | 118 | 326 | 1137.05 | 9.69 |
| 45 | Joseph Beuys | 18.7 | 11 | 194 | 3270.62 | 6.49 |
| 9 | Basquiat | 10.58 | 3 | 45 | 7124.31 | 4.07 |
| 3 | Van Gogh | 2.46 | 1 | 6 | 8640.25 | 0.98 |

**Table 3.** *The reordering of Table 2 by media index rank (descending order).*

This table is further illustrated by the graph which shows intervals of the momentum leaders from the highest-ranked Van Gogh, to the lowest-ranked, Danielle McKinney. Below we show intervals for 10 momentum leaders for the 90 days from May,1,2021 till August,1,2021. The number with the star shows the rank of the momentum leader, for example, 3* refers to *Van Gogh* (ranked 3rd by media index among all artists) and 9* to *Basquiat* (ranked 9th by media index among all artists). The interval (3,45) for Basquiat means that he has dominated "heavy hitters" ranked between 4th till 44th both in total gain as well as share gain.

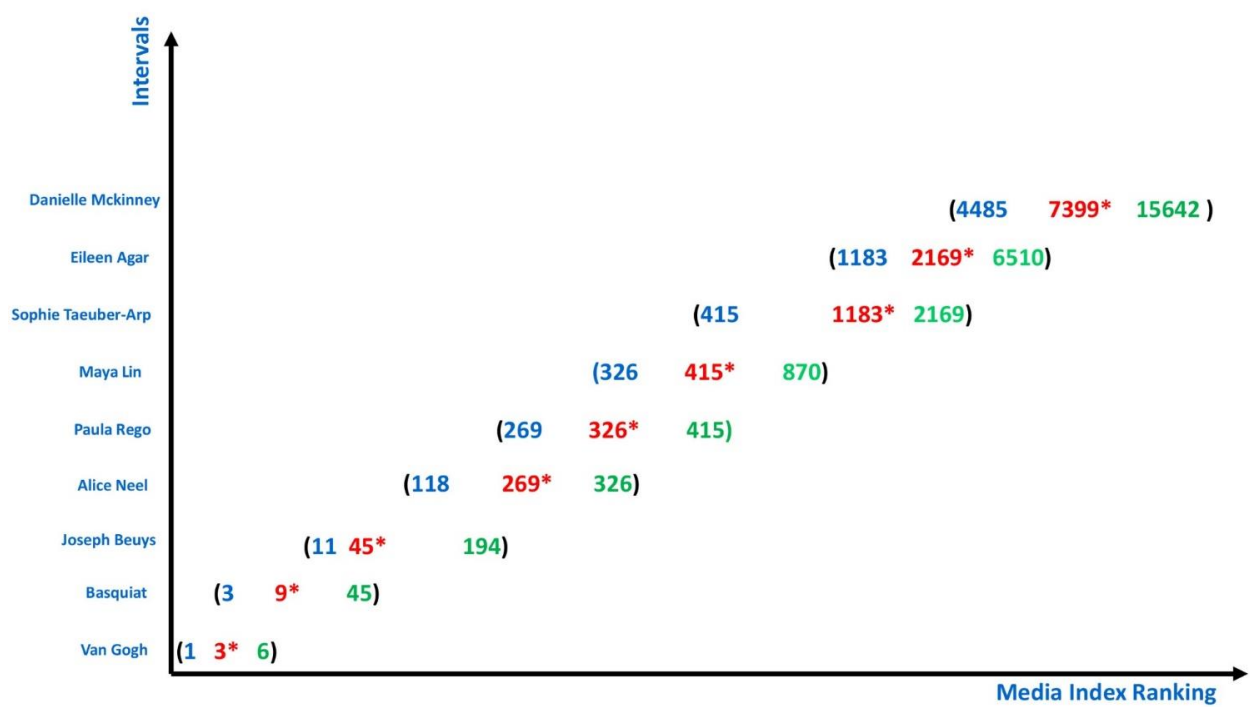

**Figure 9.** *Intervals of cohorts of artists dominated by each of the momentum leaders of August 1, 2021. Momentum leader's media index ranking in red, the lower bound of interval (ranking) in blue, and the upper bound of interval in green.*

Each interval represents a "cohort" of artists with a ranking around the artists who dominate the interval. Thus *Van Gogh's* absolute and relative gain are both higher than artists ranked between 1 and 6. *Basquiat* who is ranked number 9, had a higher relative gain than *Van Gogh*. Notice that both *Basquiat* and *Van Gogh* dominate artists ranked between them: *Banksy* (4) *DaVinci* (5) *Rembrandt* (6), *Weiwei* (7), and *Hirst* (8).

There is another very important intuition behind our notion of momentum. Momentum intervals justify why other artists have not been listed on our list of highest momentum artists. Given such a candidate artist, we can always find an artist in his/her cohort who dominate him/her in the $>>_m$ ranking, on the Pareto Frontier of $>>_m$

In other words the answer to the question "why Banksy is not a momentum leader?" is "because of van Gogh", or "because of Basquiat" (first had a higher absolute gain, the second both absolute and relative gain higher than *Banksy*.

### Why another artist is not among momentum leaders?

The proposed definition of the set of momentum leaders helps also to answer the question "why not X"? Any editorially created list of up-trending artists suffers from the subjective-ness curse – why some specific artists were omitted?

Here, we always have justification – someone is NOT on the least of momentum leaders because there is a momentum leader who dominates them.

Below, we list some examples:

a) *Why **Wangetch Mutu** is not a momentum leader?*

Because much lower-ranked Sophie Taeuber-Arp had both higher absolute gain and higher relative gain. Indeed it is 866.66 > 682.29 (total gain) and 35.62 >9.41 (relative gain).

Thus, she deserves to be listed among the top momentum artists rather than Mutu, who by the way is up-trending impressively, but not as impressive as Sophie Taeuber-Arp.

**Thus, Sophie Taeuber-Arp is the reason why Wangetch Mutu is not on the list of momentum leaders for August 1, 2021.**

b) Why ***Kandis Williams*** is not a momentum leader?

Because, Eileen Agar, had both higher relative gain and higher absolute gain than Kandis Wiliams Indeed it is 801.55 >375..63 (absolute gain) and 80.6 >59.88 (in relative gain).

**Thus, Eileen Agar is the reason why Kandis Willaims is not on the list of momentum leaders for August 1, 2021.**

c) Why ***Bene Bergado*** is not a momentum leader?

Danielle McKinney's absolute gain and relative gain (332 and 198%) dominated Bene Bergado.

**Thus, Danielle McKinney is the reason why Bene Bergado is not on the list of momentum leaders for August 1, 2021.**

Momentum Intervals cover the whole universe of artists in such a way, that for any artist, either the artist is one of the top Momentum artists (dominates one of the intervals which cover the entire universe), or is dominated by one of the top Momentum Artists. Thus for every artist, who is not a momentum leader, we justify, by showing the momentum leader who dominates such artist the Pareto $>>_m$ ranking.

### 4. Conclusions

We have demonstrated the data-driven, algorithmic method to generate the "artists to watch lists"

Our algorithm is an application of ranking by momentum method which we proposed in [1] to establish monthly lists of artists who are momentum leaders for the preceding 90 days. We use our algorithm to produce such lists on articker.org.

The lists of momentum leaders computed this way has the following properties

a) It is small (because of the small frontier property [1]).
b) It is unbiased – with no bias towards famous artists nor emerging, and yet unknown, artists
c) It is fair – artists who are not included in the list **must be dominated in terms of absolute and** relative gain by at least one member of the momentum leaders list (Pareto frontier)
d) It is objective – it is computed automatically, not editorially selected.

Ranking by momentum can be used for any subset of artists over any period of time. We may look at longer periods of possibly 6 or even 12 months, or down to 30-90 days. We may also consider subsets of emerging artists, by restricting the population of artists by age, as well as by the category of art they practice.